# SIGNIFICANCE LEVEL AND POSITIVITY BIAS AS CAUSES FOR HIGH RATE OF NON-REPRODUCIBLE SCIENTIFIC RESULTS?

JEAN-CHRISTOPHE MOURRAT

ABSTRACT. The high fraction of published results that turn out to be incorrect is a major concern of today's science. This paper contributes to the understanding of this problem in two independent directions. First, Johnson's recent claim [1] that hypothesis testing with a significance level of $\alpha = 0.05$ can alone lead to an unacceptably large proportion of false positives among all results is shown to be unfounded. Second, a way to quantify the effect of "positivity bias" (the tendency to consider only positive results as worthwhile) is introduced. We estimate the proportion of false positives among *positive* results in terms of the significance level used and the positivity ratio. The latter quantity is the fraction of positive results over all results, be they positive or not, published or not. In particular, if one uses a significance level of $\alpha = 0.05$, and produces 4 (possibly unpublished) negative results for every positive result, then the proportion of false positives among positive results can climb to a high 21%.

What are the main reasons behind the very significant proportion (see [2–4]) of scientific findings that fail to replicate? The goal of the present paper is to discuss two independent aspects of this question. In the first part, we show that Johnson's recent proposal that the use of a significance level of $\alpha = 0.05$ can alone lead to a high proportion of false positives among all results is unfounded. In short, when discussing a fraction of false positives, the numerator of this fraction (the number of false positives) is always unambiguous, but the denominator is often kept implicit, and may vary from place to place, leading to unwarranted conclusions. In the second part, we discuss a way to evaluate the ratio of false positives among *positive* results only, in terms of the significance level and the *positivity ratio*, which is defined as the proportion of positive results among all results. We will see that when one uses the significance level $\alpha = 0.05$, even moderately low values of the positivity ratio can lead to a high fraction of false positives among positive results.

## 1. Johnson's claim and refutation

Assume that we want to evaluate the adequacy of a particular null hypothesis $H_0$ against data. A standard procedure for doing so consists in specifying a significance level $\alpha \ll 1$, and constructing a set of possible outcomes that are so extreme that if $H_0$ is true, then the probability to observe a value in this set of extreme outcomes is no more than $\alpha$. Once this set of extreme outcomes (the rejection region) is defined, one performs the experiment and checks whether the result falls within this set. If so, the test is said to be positive (or significant), and the null hypothesis is rejected. The test is a false positive if it is positive although $H_0$ is actually true.

A common reformulation of this procedure goes through the computation of the so-called $p$-value. In this framework, the event that the test is positive corresponds to the fact that the $p$-value is smaller than $\alpha$.

In the long run, the proportion of false positives produced by a scientist will depend on the significance level she chooses, but also on the fraction of cases investigated where the associated null hypothesis is correct. If a scientist chooses a significance level of $\alpha = 0.05$, even in the worst-case scenario where the null hypothesis is correct in every case, she is guaranteed that the proportion of false positives among all results be no more than 5% in the long run.

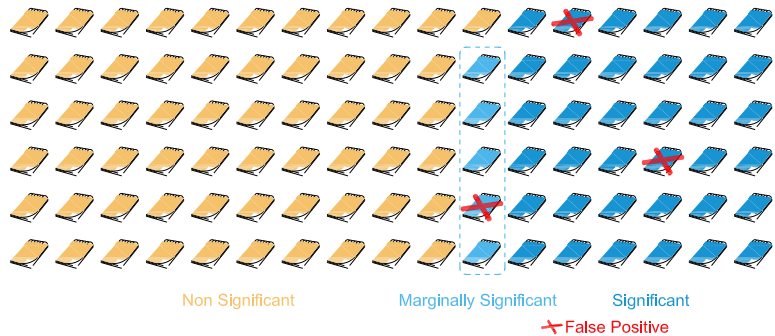

FIGURE 1. On this sample of 102 results, only 3 are false positives, but 20% of the marginally significant ones are false positives.

In Ref. [1], Johnson seems to reach a different conclusion. His main message is that the use of a significance level of $\alpha = 0.05$ can alone explain the high proportion of results that fail to reproduce (and that standard statistical practice must therefore be revised). This can be summarized as follows.

**Johnson's Claim 1.** *Fixing the significance level at $\alpha = 0.05$ can cause the proportion of false positives among all results to be unacceptably large.*

One gathers from Ref. [1] that "unacceptably large" should be understood as "about 20% or more". Johnson explains that this effect does not take publication bias into account, so "among all results" includes positive as well as negative ones.

Yet, this contradicts our initial observation on hypothesis testing. The contradiction dissipates once we notice that Johnson's argument actually shows:

**Johnson's Claim 2.** *Fixing the significance level at $\alpha = 0.05$ can cause the proportion of false positives among marginally significant results to be as high as* 20%.

The crucial difference between this and Johnson's Claim 1 is the phrase "among marginally significant results". A result is marginally significant if the observed data is extreme enough to fall within the rejection region, but just barely so. This corresponds to obtaining a $p$-value below but very close to $\alpha$.

In order to best see that Johnson's Claim 2 does not imply his Claim 1, it suffices to imagine a scenario where the null hypothesis of every test is correct. In such a case, 100% of significant results are false positives, and in particular, every marginally significant result is a false positive. Yet, in the long run, the proportion of false positives among the totality of results (be they positive or negative) is 5%.

In any case, Johnson's Claim 2 will not bring about many false positives as long as the proportion of marginally significant results remains small. A typical situation is illustrated in Figure 1, where 20% of marginally significant results are false positives, but only 3 results over 102 are false positives.

In the precise setting explored in Ref. [1], although Johnson's Claim 2 is true, it turns out that the probability to get a positive result is 0.275, while the probability for a positive result to be a false positive is about 0.09. This probability of 0.09 is markedly smaller than 20% since it is averaged over all significant results, not only over marginally significant ones. Overall, the probability of a false positive is thus $0.275 \times 0.09 = 0.025$. This means that in the long run, the proportion of false positives among all results is no more than $2.5\% \leqslant 5\%$, as expected. All this will be explained in much greater details in Supplementary Information [S1], but the reader can also directly have a look at Property 3 there, which turns Figure 1 into a more quantitative statement.

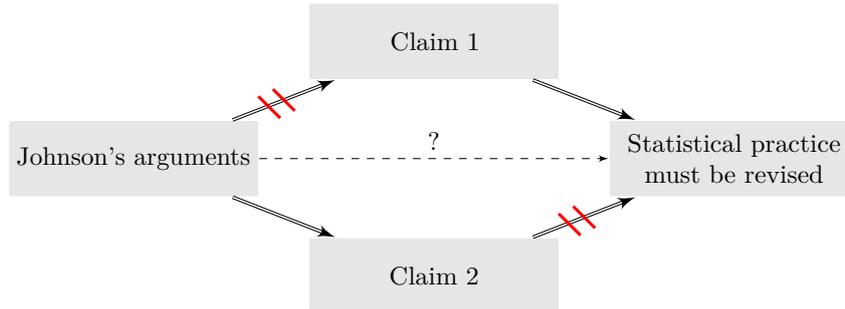

Figure 2. Claim 1 is not substantiated by Johnson's arguments; Claim 2 does not call for a revision of standard statistical practice.

To sum up, Johnson's Claim 1 is unfounded, while his Claim 2 does not call for a revision of standard statistical practice (see Figure 2). Fixing the significance level at $\alpha = 0.05$ does guarantee a proportion of false positives among all results of no more than 5%, and possibly less, depending on the fraction of cases studied where the null hypothesis is actually true.

A well-known shortcoming of hypothesis testing is that as is, a significant result gives no information about the actual size of the deviation from the null hypothesis. One possible way to move beyond this limitation is to use Bayesian reasoning, but a difficulty lies in the definition of an appropriate a priori Bayesian assumption. Another goal of Ref. [1] is to advertise the use of a priori Bayesian assumptions defined via "objective" optimization methods. Defining an a priori Bayesian assumption in such a way leads however to incoherent results, as will be discussed in Supplementary Information [S1], and is thus best avoided. In situations with no reasonably good Bayesian a priori assumption in sight, the evaluation of a confidence interval for the quantity of interest seems to be a much better practice.

## 2. Controlling the effects of positivity bias

As was reiterated above, fixing the significance level at $\alpha = 0.05$ guarantees that no more than 5% (and probably less) of obtained results are false positives. Yet, this may not be a good measure of the proportion of false positives among published papers, due to publication bias [5, 6].

In fact, publication bias is only one manifestation of a more general tendency to give more importance to positive results than to negative ones. Indeed, even in the absence of publication bias, if the readership only focuses on positive results, the net effect is identical. Let us call this tendency to give more value to positive results than to negative ones the *positivity bias*.

This bias is not necessarily a bad thing in itself. For instance, a pharmaceutical company searching for new drugs to develop will understandably focus more on results that show a therapeutic effect than not. Naturally, it will also pay some attention to negative results (e.g. by looking for studies that contradict positive studies of interest), but when assessing the proportion of studies that fail to replicate, this will be taken over a set of positive studies only.

In short, when the high rate of non-reproducible results is discussed, it is best understood as a proportion over positive results only. As was seen, the guarantee provided by standard statistical tests is different, being about the proportion of false positives over all results, be they positive or negative. The aim of this section is to discuss how the guarantee is affected by this change of viewpoint.

Without further assumptions, there is of course no way to guarantee any control on the rate of false positives among positive results. Indeed, it may be that a scientist studies only cases where her null hypothesis is actually correct, and in such a situation, all positive results are false positives.

|         | $\alpha = 0.1$ | $\alpha = 0.05$ | $\alpha = 0.01$ | $\alpha = 0.005$ |
|---------|----------------|-----------------|-----------------|------------------|
| $r = 1/2$  | 11%  | 5.3% | 1.0% | 0.5% |
| $r = 1/5$  | 44%  | 21%  | 4.0% | 2.0% |
| $r = 1/10$ | 100% | 47%  | 9.1% | 4.5% |

Table 1. Maximal proportion of false positives among positive results, as a function of the significance level $\alpha$ and the positivity ratio $r$, according to Formula (1).

One possible way to get a more meaningful result is to assume that in, say, half of the cases, the scientist's intuition is correct, i.e. the null hypothesis is wrong. This assumption has two drawbacks. First, it is not sufficient per se to rule out a situation of 100% of false positives among positive results (indeed, it could happen that all cases where the null hypothesis is wrong are missed by the scientist), so it would need to be complemented by other assumptions. Second, it is very difficult in practice to evaluate the validity of an assumption involving the proportion of cases where the null hypothesis is actually correct.

It is much more fruitful to reason in terms of the fraction of positive results over all results produced, be they positive or negative, published or kept private. Let us call this ratio the *positivity ratio $r$*. For one, each scientist can easily have a rough idea of her own positivity ratio. Secondly, one can devise a useful prediction on the fraction of false positives among *positive* results, in terms of the positivity ratio $r$ and the significance level $\alpha$ used. Indeed, the proportion of false positives among positive results can be shown (see Supplementary Information [S2]) to be no larger than

$$\text{(1)} \qquad \frac{\alpha(1-r)}{r(1-\alpha)}.$$

Table 1 gives this maximal proportion of false positives among positive results for a selection of values of $\alpha$ and $r$. For instance, for a positivity ratio of $1/2$ and for $\alpha = 0.05$, this guarantees that no more than about 5.3% of positive results are false positives. Remarkably, *for a positivity ratio of $1/5$ (which corresponds to producing 4 negative results for every positive result), statistical tests with significance level $\alpha = 0.05$ can yield a proportion of false positives among positive results as large as 21%*. In such a case, the use of a significance level of $\alpha = 0.01$ restores a proportion of false positives among positive results smaller than 5%. Similarly, for $r = 1/10$, a significance level of $\alpha = 0.005$ guarantees a proportion of false positives among positive results smaller than 5%.

A slightly different way to think about Formula (1) is displayed in Figure 3.

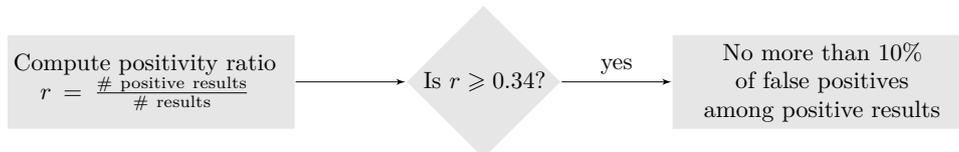

Figure 3. A simple flowchart for a significance level $\alpha = 0.05$, derived from Formula (1).

## 3. Conclusions

First, we re-affirmed the fundamental properties of hypothesis testing, and showed that Johnson's claim [1] that choosing a significance level of $\alpha = 0.05$ can alone lead to a very high proportion of false positives among all results is unfounded.

Second, we explored the interactions between the choice of the significance level and positivity bias. We introduced the positivity ratio $r$, which is the fraction of positive results over the totality of results, be they positive or not, published or not. One can bound the proportion of false positives among *positive* results in terms of the positivity ratio and the significance level $\alpha$ used, according to Formula (1). This proportion can be as high as 21% if one assumes $r = 1/5$ and $\alpha = 0.05$.

It is natural to wonder how these observations can be turned into recommendations. These should depend on what the positivity ratio of the scientific community as a whole (or within subfields) is. If it is about $1/3$ or more, then the common practice of using a significance level of $\alpha = 0.05$ guarantees a small ($\leqslant 10\%$) proportion of false positives among positive results. If instead, the positivity ratio is observed to be about $1/5$ or less, then one should seriously consider moving to a significance level of $\alpha = 0.01$, so as to bring the high 21% or more of predicted false positives among positive results back to a more reasonable fraction. Using a significance level of $\alpha = 0.01$ would guarantee a proportion of false positives among positive results of no more than 5% as soon as $r \geqslant 0.17$, and of no more than 10% as soon as $r \geqslant 0.092$. In most cases, scientific endeavors for which data collection is costly or time-consuming can probably not afford smaller positivity ratios. On the other hand, the statistical treatment of data-intensive subjects must be handled with more precaution.

A final note of caution is in order. So far, we have only considered problems that could arise from insufficiently strong statistical tests and positivity bias. An analysis of retracted biomedical and life-science research articles [7] has however revealed that a large majority of retractions were in fact attributable to scientific misconduct. This problem, of course, was not taken into account in this paper.

**Acknowledgements.** I would like to thank Erwin Berthier and Yoav Zemel for stimulating discussions, and Rosanne Berthier for designing Figure 1.

## References

[S1] **Supplementary Information S1**. *On Johnson's arguments.* The claims made in the main text that relate to Johnson's arguments are substantiated in detail.

[S2] **Supplementary Information S2**. *Proof of Formula (1).* The proportion of false positives among positive results is shown to be no more than the fraction given in (1) of the main text.

[1] V.E. Johnson. Revised standards for statistical evidence. *Proc. Nat. Acad. Sci. USA* **110** (48), 19313-19317 (November 26, 2013).

[2] J.P.A. Ioannidis. Why most published research findings are false. *PLoS Med.* **2** (8), e124 (2005).

[3] F. Prinz, T. Schlange, K. Asadullah. Believe it or not: how much can we rely on published data on potential drug targets? *Nat. Rev. Drug Discov.* **10**, 712 (2011).

[4] C.G. Begley, L.M. Ellis. Drug development: Raise standards for preclinical cancer research. *Nature* **483**, 531–533 (2012).

[5] P.J. Easterbrook, J.A. Berlin, R. Gopalan, D.R. Matthews. Publication bias in clinical research. *Lancet* **337** (8746), 867–872 (1991).

[6] F. Song, S. Parekh, L. Hooper, Y.K. Loke, J. Ryder, A.J. Sutton, C. Hing, C.S. Kwok, C. Pang, I. Harvey. Dissemination and publication of research findings: An updated review of related biases. *Health Technol. Assess.* **14** (8) (2010).

[7] F.C. Fang, R.G. Steen, A. Casadevall. Misconduct accounts for the majority of retracted scientific publications. *Proc. Nat. Acad. Sci. USA* **109** (42), 17028-17033 (October 16, 2012).

ENS Lyon, CNRS, 46 allée d'Italie, 69007 Lyon, France

# ON JOHNSON'S ARGUMENTS

The aim of this supplementary information is to substantiate the claims made in the main text that relate to Johnson's arguments [1]. It is essentially self-contained, so that I hope that any interested scientist can read it.

In Section 1, we review the main features of hypothesis testing. Johnson's argument is based on Bayesian reasoning. The aim of Section 2 is to introduce such a setting and show an apparently paradoxical situation that is similar, but simpler, than that encountered by Johnson. In Section 3, we present Johnson's setting and assumptions proper, and explain why they do not imply the announced high rate of false positives among all results. The goal of Section 4 is to show that Johnson's proposed method for defining an a priori Bayesian hypothesis leads to incoherent results, and should thus be avoided. Section 5 backs the contentious claims attributed to Johnson with quotations from Ref. [1].

## 1. Hypothesis testing

We begin by describing an example of hypothesis testing that is best suited for a comparison with Johnson's setting. Let us imagine that Alice plans to run an experiment that consists in repeating $n$ times independently the same protocol, whose aim is to measure a (dimensionless) quantity of interest. In the end of the experiment, she will have gathered $n$ measurements $(x_1, \ldots, x_n) = \mathbf{x}$. Her *null hypothesis* is that the distribution of these measurements is Gaussian with mean 0 and variance 1. Her scientific motivation for the experiment is that she suspects that the mean may in fact be positive. Her plan is to compute the empirical mean $\bar{x} = (x_1 + \cdots + x_n)/n$, and hopefully observe that it is significantly above 0.

In order to give this idea a precise sense, she decides to devise a hypothesis test with significance level $\alpha = 0.05$. Under the null hypothesis, the empirical mean $\bar{x}$ is centered Gaussian with mean 0 and variance $1/n$, so the probability that $\bar{x}$ exceeds $1.65/\sqrt{n}$ is about $0.05$.[1] In view of this, she decides that she will have disproved the null hypothesis if at the end of her experiment, she finds $\bar{x} \geqslant 1.65/\sqrt{n}$. She has ensured the following property.

**Property 1.** *If the null hypothesis $H_0$ is true, then the probability that the test indicates to reject the null hypothesis (a case of false positive) is no more than $\alpha = 0.05$.*

The statistical test designed by Alice is just one particular example of hypothesis testing with significance level $\alpha = 0.05$. In any case, we want Property 1 to be satisfied. By the law of large numbers, Property 1 implies the following.

**Consequence 1.** *Assume that Alice performs $K \gg 1$ experiments, each with a significance level of $\alpha = 0.05$. Assume also that the null hypothesis is true in half of her experiments. Then the proportion of false positives among all results is no more than $2.5\%$ (i.e. $5\%$ of the cases where the null hypothesis is true).*

Under the same assumptions but with a different reasoning, Johnson [1] reaches the conclusion that the proportion of false positives should rather be of 20%, or possibly more. This is in contradiction with Consequence 1, and our purpose is to see why Johnson's claim is unfounded.

---

[1] The mathematically minded reader may prefer to replace the approximate value 1.65 by $x^*$ such that $\frac{1}{\sqrt{2\pi}} \int_{x^*}^{+\infty} e^{-x^2/2} \, dx = 0.05$.

One could rephrase Alice's hypothesis test in terms of the computation of the $p$-value. I will not recall what the $p$-value is here, but simply mention that the test as described above corresponds to checking whether the $p$-value is below $\alpha = 0.05$.

## 2. Bayesian reasoning and an apparent paradox

The goal of this section is to explain in a simple setting some apparently paradoxical facts that I believe to be similar to those encountered by Johnson.

We want to reconsider Alice's problem using a Bayesian approach. The starting point of Bayesian reasoning is to define *a priori* probabilities on the set of possible distributions of Alice's quantity of interest. Let us suppose that Alice chooses to encode her a priori belief in the results of her experiment as follows. Her a priori assumption is that for some specific $\mu > 0$ of her choosing,

(BH) $\quad$ with probability $1/2$, the hypothesis $H_0$ is true,
$\quad\quad\;\;$ with probability $1/2$, the hypothesis $H_\mu$ is true,

where $H_0$ (resp. $H_\mu$) is the hypothesis that the measurements are distributed as independent Gaussian random variables of mean 0 (resp. $\mu$) and variance 1.

This assumption could be justified for instance if Alice collects her data from Bob, and knows that he generates them in the following way. Bob (secretly) tosses a fair coin. If he gets "head", then he supplies Alice with independent Gaussian variables of mean 0 and variance 1; otherwise, he supplies her with independent Gaussian variables of mean $\mu$ and variance 1. Of course, this example is only meant to guide our intuition, but more realistic scenarios can be constructed. For instance, Alice studies a population, and has noticed that she can subdivide it into two subgroups $G_1$ and $G_2$ of equal size (based on, say, whether the individual carries a specific gene or not). She also noticed that if she repeats some measurement on an individual in $G_1$ (resp. $G_2$), she gets independent Gaussian variables of mean 0 (resp. $\mu$) and variance 1. She is then interested in an individual drawn from the entire population, and wonders whether it belongs to $G_1$ or $G_2$. Both settings fit perfectly with assumption (BH). In the latter case, the event that the individual belong to $G_1$ (resp. $G_2$) corresponds to hypothesis $H_0$ (resp. $H_\mu$).

It certainly comes to the reader's mind that in many cases of interest, there will be no way for Alice to come up with a specific value of $\mu$ such that assuming (BH) becomes reasonable[2]. We will return to this critical question below, and focus for now on exploring the mathematical implications of this assumption.

Under $H_0$ (resp. $H_\mu$), the empirical mean $\overline{x}$ follows a Gaussian distribution with mean 0 (resp. $\mu$) and variance $1/n$. Hence, under the Bayesian hypothesis (BH), the empirical mean is distributed according to the probability density

$$\frac{1}{2}\frac{\sqrt{n}}{\sqrt{2\pi}}\left(e^{-n\overline{x}^2/2} + e^{-n(\overline{x}-\mu)^2/2}\right)\,\mathrm{d}\overline{x}.$$

The Bayesian method consists in computing (using Bayes formula) the a posteriori probabilities of $H_0$ and $H_\mu$ given the data $\mathbf{x}$ that was observed:

$$P[H_0 \mid \mathbf{x}] = \frac{p(\mathbf{x} \mid H_0)P[H_0]}{p(\mathbf{x})},$$

where $p(\mathbf{x})$ is the probability density of the variable $\mathbf{x}$ (assuming (BH)), and $p(\mathbf{x} \mid H_0)$ is the probability density of $\mathbf{x}$ conditionally on $H_0$. A computation shows that this is equal to

$$\frac{p(\overline{x} \mid H_0)P[H_0]}{p(\overline{x})} = \frac{e^{-n\overline{x}^2/2}}{e^{-n\overline{x}^2/2} + e^{-n(\overline{x}-\mu)^2/2}},$$

---

[2]The choice of the probabilities fixed at $1/2 - 1/2$ in (BH) is another problem that we will leave aside for simplicity.

which can be rewritten as

$$P[H_0 \mid \mathbf{x}] = \frac{1}{1 + \mathsf{BF}(\mathbf{x})}, \tag{1}$$

where $\mathsf{BF}(\mathbf{x})$ is the *Bayes factor*

$$\mathsf{BF}(\mathbf{x}) = e^{n(2\mu\overline{x} - \mu^2)/2}. \tag{2}$$

Similarly, we get[3]

$$P[H_\mu \mid \mathbf{x}] = \frac{1}{1 + 1/\mathsf{BF}(\mathbf{x})}. \tag{3}$$

Hence, a large Bayes factor indicates that $H_\mu$ is more likely to be true, while a small one gives support for $H_0$.

In view of assumption (BH), if $n$ is very large, we have

$$\overline{x} \simeq \left| \begin{array}{ll} 0 & \text{with probability } \simeq 1/2, \\ \mu & \text{with probability } \simeq 1/2, \end{array} \right. \tag{4}$$

and as a consequence, the distribution of the Bayes factor is such that[4]

$$\mathsf{BF}(\mathbf{x}) \simeq \left| \begin{array}{ll} e^{-n\mu^2/2} & \text{with probability } \simeq 1/2, \\ e^{n\mu^2/2} & \text{with probability } \simeq 1/2. \end{array} \right. \tag{5}$$

In particular, when $n$ is very large, *it is extremely unlikely that the Bayes factor takes moderate values*, as for instance being in the interval $[1/20, 20]$. Hence, a test that would consist in declaring that the hypothesis $H_\mu$ is empirically proved when $\mathsf{BF}(\mathbf{x}) \geqslant 1$ *would actually yield a false positive with very small probability* if $n$ is very large, assuming that our a priori assumption (BH) is correct. This can look paradoxical, since the information $\mathsf{BF}(\mathbf{x}) \geqslant 1$ seems basically useless at first sight. Indeed, by (1), it only indicates that the a posteriori probability for $H_0$ is smaller than $1/2$. Yet, for large $n$, the assumption (BH) forces the Bayes factor to have the distribution described in (5), and thus outside of an event of very small probability, asking to have $\mathsf{BF}(\mathbf{x}) \geqslant 1$ in fact forces the Bayes factor to be about $e^{n\mu^2/2}$, and thus $P[H_0 \mid \mathbf{x}] \simeq e^{-n\mu^2/2}$. Let us stress this:

**Property 2.** *If Alice is right in her a priori assumption* (BH), *and if $n$ is very large, then the following holds. Conditionally on the event $\mathsf{BF}(\mathbf{x}) \geqslant 1$, although we can only know for sure that $P[H_0 \mid \mathbf{x}] \leqslant 1/2$, outside of a very unlikely event, we actually have $\mathsf{BF}(\mathbf{x}) \simeq e^{n\mu^2/2}$ and $P[H_0 \mid \mathbf{x}] \simeq e^{-n\mu^2/2}$. In particular,*

$$P[H_0 \mid \mathsf{BF}(\mathbf{x}) \geqslant 1] \ll 1.$$

**Consequence 2.** *If Alice is right in her a priori assumption* (BH), *and if $n$ is very large, then deciding to reject the null hypothesis $H_0$ when $\mathsf{BF}(\mathbf{x}) \geqslant 1$ yields a very small probability of false positive (i.e. a case where Alice rejects $H_0$ although it was correct).*

---

[3]and recover the classical definition of the Bayes factor,
$$\mathsf{BF}(\mathbf{x}) = \frac{P[H_\mu \mid \mathbf{x}]}{P[H_0 \mid \mathbf{x}]} \frac{P[H_0]}{P[H_\mu]} = \frac{P[\mathbf{x} \mid H_\mu]}{P[\mathbf{x} \mid H_0]}.$$

[4]My use of the sign $\simeq$ is ambiguous, but the interested reader will easily make the statements (4) and (5) more precise. For instance, we can pick some $\varepsilon > 0$ very small (in particular, smaller than $\mu/2$), and replace the statement "$\overline{x} \simeq 0$ with probability $\simeq 1/2$" by "$P[-\varepsilon \leqslant \overline{x} \leqslant \varepsilon] \xrightarrow[n \to \infty]{} 1/2$", while replacing the statement "$\mathsf{BF}(\mathbf{x}) \simeq e^{-n\mu^2/2}$ with probability $\simeq 1/2$" by
$$P\left[-\frac{\mu^2}{2} - \varepsilon\mu \leqslant \frac{\log(\mathsf{BF}(\mathbf{x}))}{n} \leqslant -\frac{\mu^2}{2} + \varepsilon\mu\right] \xrightarrow[n \to \infty]{} 1/2.$$

This can be made precise. For instance, one can check that as soon as $\sqrt{n} \geqslant 2 \times 1.65/\mu$, the test $\mathsf{BF}(\mathbf{x}) \geqslant 1$ to reject $H_0$ has a probability of yielding a false positive smaller than 0.025.

This being said, and although the policy described in Consequence 2 leads to a very small probability of false positive when $n$ is large, it would be strange for Alice to choose the criterion $\mathsf{BF}(\mathbf{x}) \geqslant 1$ as her condition for when to reject $H_0$. Indeed, on the event that $\mathsf{BF}(\mathbf{x}) \geqslant 1$ but $\mathsf{BF}(\mathbf{x}) \simeq 1$, the evidence in favor of the alternative hypothesis $H_\mu$ is only very weak. In fact, Consequence 2 would remain true if the condition $\mathsf{BF}(\mathbf{x}) \geqslant 1$ was replaced by, for instance, the condition $\mathsf{BF}(\mathbf{x}) \geqslant 1/20$. Yet, in the case when $1/20 \leqslant \mathsf{BF}(\mathbf{x}) \leqslant 1$, the evidence actually favors hypothesis $H_0$, so it would be foolish to reject it in such a case. There is no contradiction however, since the event that $1/20 \leqslant \mathsf{BF}(\mathbf{x}) \leqslant 1$ has negligible probability when $n$ is very large, so that in practice, Alice will almost never encounter such a situation.

It is important to notice that all our reasoning relies on our assuming (BH) to be a valid a priori assumption.

## 3. Johnson's setting

In the case when Alice has no good prior information such as (BH) at hand, Johnson argues that she could start with the a priori assumption (BH) with $\mu$ given by a sort of automatic procedure that makes it the "best" possible $\mu$ in the following sense. Alice gives herself a parameter $\gamma$ and decides that she will support the hypothesis $H_\mu$ as soon as she finds that the Bayes factor $\mathsf{BF}(\mathbf{x})$ exceeds $\gamma$. Then $\mu$ is chosen so as to maximize Alice's chances to find support for the alternative hypothesis $H_\mu$, that is, to find $\mathsf{BF}(\mathbf{x}) \geqslant \gamma$. Johnson finds that this optimal $\mu$ is[5]

$$\mu = \mu(\gamma, n) = \sqrt{\frac{2 \log(\gamma)}{n}}. \tag{6}$$

With this choice of $\mu$, a computation leads him to the observation that for $\gamma = 3.87$, the condition[6]

$$\{\mathsf{BF}(\mathbf{x}) \geqslant \gamma\}$$

is precisely the same as the condition

$$\{\overline{x} \geqslant 1.65/\sqrt{n}\},$$

that is, the condition obtained in Section 1 for rejecting the null hypothesis with significance level $\alpha = 0.05$. He then concludes that this gives only poor support for assumption $H_\mu$, since the condition $\mathsf{BF}(\mathbf{x}) \geqslant \gamma$ with $\gamma = 3.87$ only means that the a posteriori probability of $H_\mu$ is about 0.8, see (1).

The apparent contradiction is similar to that found in Property 2 and Consequence 2. Indeed, similarly to what we saw there, under assumption (BH) and on the event that $\mathsf{BF}(\mathbf{x}) \geqslant 3.87$, the Bayes factor is likely to be significantly larger than the value 3.87. The effect is not as drastic as in Property 2 however. Indeed, on the event $\mathsf{BF}(\mathbf{x}) \geqslant 3.87$ (and assuming (BH) to be correct), the Bayes factor will have some small but non-negligible probability of taking a value that is only moderately large, some other probability to be larger, yet some other probability of being very large, and so on. Each of these sub-cases gives rise to a certain conditional probability of a false positive. However, the total probability of a false positive is an average over all these sub-cases, and by design, it remains smaller than 0.025. In formulas,

$$P[\text{false positive}] = P[\mathsf{BF}(\mathbf{x}) \geqslant \gamma \text{ and } H_0] = \frac{1}{2} \, P[\mathsf{BF}(\mathbf{x}) \geqslant \gamma \mid H_0],$$

---

[5]log denotes the natural logarithm

[6]The mathematically minded reader may prefer to replace 3.87 by $\gamma^* = e^{(x^*)^2/2}$, where $x^* \simeq 1.65$ was defined in footnote (1).

| BF(**x**) | 3.87 | 5.44 | 7.92 | 12.31 | 21.77 | $\infty$ |
|---|---|---|---|---|---|---|
| $p$-value | 0.05 | 0.032 | 0.019 | 0.0094 | 0.0035 | 0 |
| $P[E]$ | 0.05 | 0.05 | 0.05 | 0.05 | 0.075 | |
| $P[H_0\|E]$ | 0.18 | 0.13 | 0.09 | 0.06 | 0.03 | |

TABLE 1. The first line of the table describes possible values of the Bayes factor (and the corresponding $p$-value on the second line). On the last two lines, a column in between two possible values of the Bayes factor refers to the event $E$ that the Bayes factor falls in between these two values, see Property 3.

and for $\gamma = 3.87$, since $\{\mathsf{BF}(\mathbf{x}) \geqslant \gamma\} = \{\overline{x} \geqslant 1.65/\sqrt{n}\}$,

$$P[\mathsf{BF}(\mathbf{x}) \geqslant \gamma \mid H_0] = P[\overline{x} \geqslant 1.65/\sqrt{n} \mid H_0] = 0.05,$$

so that indeed $P[\text{false positive}] = 0.025$. We illustrate this idea further in Table 1 by computing, for a selection of intervals, the probability that the Bayes factor falls within this interval, and the probability of a false positive given that the Bayes factor falls within this interval. The latter probability can be computed using (1):

$$\begin{aligned}
P[H_0 \mid \mathsf{BF}(\mathbf{x}) \in I] &= E\left[P[H_0 \mid \mathbf{x}] \mid \mathsf{BF}(\mathbf{x}) \in I\right] \\
&= E\left[\frac{1}{1 + \mathsf{BF}(\mathbf{x})} \mid \mathsf{BF}(\mathbf{x}) \in I\right],
\end{aligned} \tag{7}$$

and in particular, the observation that the total probability of a false positive is 0.025 can be rewritten using (7) as

$$P[\text{false positive}] = P[H_0 \text{ and } \mathsf{BF}(\mathbf{x}) \geqslant 3.87]$$
$$= E\left[\frac{1}{1 + \mathsf{BF}(\mathbf{x})} \mid \mathsf{BF}(\mathbf{x}) \geqslant 3.87\right] \ P[\mathsf{BF}(\mathbf{x}) \geqslant 3.87] = 0.025.$$

From the values given in Table 1, we can verify that

$$P[H_0 \text{ and } \mathsf{BF}(\mathbf{x}) \geqslant 3.87] = 0.18 \times 0.05 + \cdots + 0.03 \times 0.075 = 0.025.$$

A convenient way to interpret Table 1 and this relation is as follows.

**Property 3.** *Assume that Alice performs $K \gg 1$ experiments, each of them satisfying the a priori Bayesian assumption* (BH) *with $\mu$ given by* (6)*. Then about $0.05K$ of her results will have a Bayes factor between $3.87$ and $5.44$ (and a $p$-value between $0.05$ and $0.032$), and among these, about $0.18 \times 0.05K$ will have wrongly rejected the null hypothesis $H_0$. About $0.05K$ of her results results will have a Bayes factor between $5.44$ and $7.92$ (and a $p$-value between $0.032$ and $0.019$), out of which about $0.13 \times 0.05K$ will have wrongly rejected $H_0$, and so on, following the numbers given in Table 1. In total, she has no more than about $0.025K$ papers wrongly rejecting the null hypothesis $H_0$, as expected.*

## 4. Devising a priori hypotheses

In the previous section, we assumed that hypothesis (BH) with $\mu$ given by (6) was correct. Under this assumption, we explained the workings of the Bayesian approach, and showed why contrary to what was claimed in Ref. [1], it does not lead to a high probability of false positive.

The key point to be stressed now is that all the reasoning in the previous section can possibly make sense *only if the a priori assumption* (BH) *with $\mu$ given by* (6) *is believed to be correct*. Otherwise, even the definition of the Bayes factor is groundless. Yet, the single fact that $\mu$ was determined by the "objective" formula (6) is of course *no reason* for putting faith in assumption (BH) with this specific $\mu$. Actually, the fact that the a priori assumption (BH) is defined without any concern for the specifics of the experiment under consideration is rather a very worrying sign for the validity of (BH). Indeed, it may well be that the true distribution of Alice's measurements is *very far* from being close to any of the two distributions that appear in the a priori assumption (BH). Johnson rightly points out that an advantage of the Bayesian approach is that it can not only rule out the null hypothesis, but can also give support for the alternative hypothesis. However, as long as Alice cannot convince her readership that assuming (BH) with $\mu$ as in (6) is a valid assumption, this is not helpful at all. In order to best illustrate this, imagine the following scenarios.

(1) Alice chooses the threshold value $\gamma = 3.87$, and does $n = 100$ measurements. The empirical mean she obtains is $\overline{x} = 100$. If she follows Johnson's proposal, she assumes (BH) with $\mu = \sqrt{2\log(\gamma)/n} = 0.165$. She thus finds extremely strong support for the hypothesis that her measurements have mean 0.165, with the stratospheric Bayes factor of $10^{714}$, although recall that the empirical mean she found is 100!

(2) Alice fixes her threshold value $\gamma = 3.87$ and takes $\mu$ as in (6). Her data then shows strong support for the hypothesis $H_\mu$. Meanwhile, her colleague Bob chooses a different threshold, e.g. $\gamma = 20$, and also finds strong support for his alternative hypothesis. This is very likely to happen for certain choices of the underlying distribution, and yet note that Alice's and Bob's evidence-based alternative hypotheses are different, since they correspond to different values of $\mu$!

(3) Bob chooses the same $\gamma$ as Alice, and both independently find support for the alternative hypothesis. They then find out that they have done the same experiment. They decide to group their data, and discover that since they now have $2n$ measurements instead of $n$ each, their grouped data actually supports a different hypothesis than that of each separately!

These examples vividly illustrate that the choice of the a priori Bayesian hypothesis is not innocent. It needs to be carefully substantiated, instead of drawn from some blind (albeit "objective") automatic procedure.

As a side remark, let us note that if Alice is interested in estimating the mean of her measurements and does not have much a priori knowledge on her experiment, she can use the frequentist approach with good effect. Indeed, let us suppose that she is willing to accept the assumption that her measurements follow a Gaussian distribution with some unspecified mean $\mu$ and unspecified variance that is at most 1, and is right in doing so. This is a much weaker assumption than (BH), since in particular it does not specify a distribution on the possible values of $\mu$ themselves. Under this sole assumption, she obtains that outside of an event whose probability is smaller than 0.05, the true mean $\mu$ satisfies

$$|\mu - \overline{x}| \leqslant \frac{1.96}{\sqrt{n}}.$$

## 5. Quotations from Ref. [1]

The aim of this section is to back the contentious claims attributed to Johnson by quotations from Ref. [1].

His attributed main message is that the standard practice of fixing the significance level at $\alpha = 0.05$ can alone explain the high proportion of scientific results that fail to reproduce. This was rephrased in the main text as

**Claim 1.** *Fixing the significance level at $\alpha = 0.05$ can cause the proportion of false positives among all results to be unacceptably large.*

In the introduction of Ref. [1], we read

> "recent concerns over the lack of reproducibility of scientific studies can be attributed largely to the conduct of significance tests at unjustifiably high levels of significance."

A highlighted piece of text reads

> "a root cause of nonreproducibility is traced to the conduct of significance tests at inappropriately high levels of significance."

(We learn elsewhere that $\alpha = 0.05$ is an inappropriately high level of significance.) We have seen (and this is very clearly illustrated in Figures 1 and 2 of the main text) that Johnson's Claim 2 (cf. main text) is *not* sufficient on its own to lead to such conclusions. So I can see no way to interpret the above quotations other than as meaning that Claim 1 holds. The fact that the proportion of false positives should be taken among all results, be they positive or not, is substantiated by Johnson explicitly stating that his argument does not involve publication bias:

> "it is important to note that this high rate of nonreproducibility is not the result of scientific misconduct, publication bias, file drawer biases, or flawed statistical designs; it is simply the consequence of using evidence thresholds that do not represent sufficiently strong evidence in favor of hypothesized effects."

The confusion between Claims 1 and 2 can be seen for instance in the following passage.

> "these results suggest that between 17% and 25% of marginally significant scientific findings are false. This range of false positives is consistent with nonreproducibility rates reported by others (e.g., ref. 5). If the proportion of true null hypotheses is greater than one-half, then the proportion of false positives reported in the scientific literature, and thus the proportion of scientific studies that would fail to replicate, is even higher."

While the first sentence refers to Claim 2, the rest refers to Claim 1.

It is also worth noting that under Johnson's assumptions, the proportion of false positives among positive results is about 9% (in particular, much smaller than 20%). As a consequence, a modified Claim 1 that would involve the proportion of false positives among positive results only would not be supported by Johnson's example. Besides, outside of Johnson's assumptions, it is very easy to think of situations where *all* positive results are false positives (e.g. when all tested null hypotheses are true), no matter how small the significance level is.

A number of research papers have now appeared (see e.g. [2–9]) that refer to Johnson's article as proof that setting a significance level of 0.05 is bound to produce unreliable results. A recent editorial [10] of the journal *Toxicological Sciences* captures the essence of Johnson's message very clearly:

> "Dr Valen Johnson wrote a provocative article in the Proceedings of the National Academy of Sciences arguing that our current standard of $p < 0.05$ for significance is a major cause of scientific irreproducibility (Johnson, 2013). He proposes that science should move toward significance levels of 0.005–0.001, and he is not talking about big data science. He is talking about the routine sorts of assays scientists, including toxicologists, have used for decades. While such levels of significance take additional resources by requiring more replicates, more samples, and more animals, he makes an excellent case that there would be an overall cost savings to the scientific enterprise. Inherent to his premise is that the cost of achieving higher levels of significance would be offset by the resources wasted chasing down spurious findings."

The second contentious claim attributed to Johnson here is his recommendation of using his method of devising a priori Bayesian assumptions (in our example, to assume (BH) with $\mu$ given by (6)) for practical purposes. We read:

> "defining a Bayes factor requires the specification of both a null hypothesis and an alternative hypothesis, and in many circumstances there is no objective mechanism for defining an alternative hypothesis. The definition of the alternative hypothesis therefore involves an element of subjectivity, and it is for this reason that scientists generally eschew the Bayesian approach toward hypothesis testing. Efforts to remove this hurdle continue, however [...]. Recently, Johnson (21) proposed a new method for specifying alternative hypotheses."

> "These observations [...] suggest a simple strategy for improving the replicability of scientific research. This strategy includes the following steps: [...] When UMPBTs can be defined (or when other default Bayesian procedures are available), report the Bayes factor in favor of the alternative hypothesis and the default alternative hypothesis that was tested."

(The "default alternative hypothesis" coming from a "UMPBT" is that obtained from Johnson's procedure.)

## References

[1] V.E. Johnson. Revised standards for statistical evidence. *Proc. Nat. Acad. Sci. USA* **110** (48), 19313-19317 (November 26, 2013).

[2] E. Theodoratou et al. Vitamin D and multiple health outcomes: umbrella review of systematic reviews and meta-analyses of observational studies and randomised trials. *BMJ* **348**, g2035 (2014).

[3] D. M'boule et al. Salinity dependent hydrogen isotope fractionation in alkenones produced by coastal and open ocean haptophyte algae. *Geochim. Cosmochim. Acta* **130**, 126–135 (2014).

[4] Y. Chen et al. An MR-Conditional High-Torque Pneumatic Stepper Motor for MRI-Guided and Robot-Assisted Intervention. *Ann. Biomed. Eng.* **42** (9), 1823-1833 (2014).

[5] C. Scharinger et al. Platelet Serotonin Transporter Function Predicts Default-Mode Network Activity. *PLOS ONE* **9** (3), e92543 (2014).

[6] R. Whelan et al. Neuropsychosocial profiles of current and future adolescent alcohol misusers. *Nature* **512**, 185–189 (14 August 2014).

[7] M.P. Pavlou et al. Integrating Meta-Analysis of Microarray Data and Targeted Proteomics for Biomarker Identification: Application in Breast Cancer. *J. Proteome Res.* **13** (6), 2897–2909 (2014).

[8] C. Poland et al. The Elephant in the Room: Reproducibility in Toxicology. *Part. Fibre Toxicol.* **11** (42) (2014).

[9] R.P. Granacher Jr. Commentary: Dissociative Amnesia and the Future of Forensic Psychiatric Assessment. *J. Am. Acad. Psychiatry Law* **42** (2), 214-218 (2014).

[10] G.W. Miller. Editorial – Improving Reproducibility in Toxicology. *Toxicol. Sci.* **139** (1), 001-003 (May 2014).

# PROOF OF FORMULA (1)

The goal is to understand the fraction of false positives produced by a scientist among all her *positive* results only, under the condition that she only uses statistical tests with a fixed significance level $\alpha$. Assume that she performs $K \gg 1$ different experiments, out of which $K_+$ are positive (i.e. reached the conclusion to reject the null hypothesis). Her *positivity ratio $r$* is the fraction of positive results over the total, that is, $r = K_+/K$.

Let $\eta$ be the fraction of cases for which the null hypothesis was correct. Among the cases where the null hypothesis was wrong, let $\beta$ be the fraction of results that were negative ($\beta$ is close to 0 if the scientist is efficient in detecting when the null hypothesis is wrong). The decomposition of all possible cases, together with their relative proportions, is summarized in Table 1. The fact that out of $\eta K$ cases where the null hypothesis is true, only $\eta \alpha K$ yield a false positive, comes from the defining property of the significance level (see Property 1 of Supplementary Information S1).

|         | positive test | negative test |
|---------|---------------|---------------|
| $H_0$ true | false positive proportion $\eta\alpha$ | true negative proportion $\eta(1-\alpha)$ |
| $H_0$ false | true positive proportion $(1-\eta)(1-\beta)$ | false negative proportion $(1-\eta)\beta$ |

TABLE 1. A decomposition of possible cases, with their relative proportions.

An examination of Table 1 shows that
$$r = \eta\alpha + (1-\eta)(1-\beta) \leqslant \eta\alpha + 1 - \eta,$$
which yields
$$\eta \leqslant \frac{1-r}{1-\alpha}.$$
Hence, the proportion of false positives over the totality of positive results is at most
$$\frac{\eta\alpha}{r} \leqslant \frac{\alpha(1-r)}{r(1-\alpha)},$$
as announced.